\documentclass[fleqn,usenatbib]{mnras}

\usepackage{newtxtext,newtxmath}
\usepackage[T1]{fontenc}
\usepackage{ae,aecompl}
\usepackage{graphicx}	% Including figure files
\usepackage{amsmath}	% Advanced maths commands
\usepackage{dblfloatfix}% To enable figures at the bottom of page
\usepackage{color,soul}
\usepackage{indentfirst}
\usepackage{url}
\usepackage{scalerel}
\usepackage{placeins}
\usepackage{subfigure}
\usepackage[textwidth=1.5cm,shadow]{todonotes}	% \todo{}: side notes at the margins; \todo[inline]{}: inline notes; \missingfigure{}; \listoftodos
\usepackage[normalem]{ulem}    
\usepackage{wrapfig}	% wrap text around figures; https://en.wikibooks.org/wiki/LaTeX/Floats,_Figures_and_Captions
\usepackage{verbatim}   % "comment" env.
\usepackage{mathtools}
\newcommand{\eg}[1]{(e.g. \citealt{#1})}

\newcommand{\orcid}[1]{\href{#1}{\includegraphics[scale=0.04]{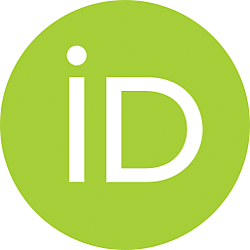}}}

\newcommand{\medd}{\dot{M}_{\rm Edd}}
\newcommand{\msun}{M_{\odot}}
\def\quote#1{``#1''}

\title[LLAGN Winds and Star Formation]{Quenching star formation with low-luminosity AGN winds}
\author[Almeida et al ]{
Ivan Almeida$^{1,2}$\thanks{E-mail: ivan.almeida003@gmail.com}\orcid{https://orcid.org/0000-0001-6018-2852},
Rodrigo Nemmen$^{1,3}$\orcid{https://orcid.org/0000-0003-3956-0331} and 
Rogemar A. Riffel$^{4}$\orcid{https://orcid.org/0000-0003-0483-3723}
\\
$^1$Instituto de Astronomia, Geof\'{\i}sica e Ci\^encias Atmosf\'ericas, Universidade de S\~ao Paulo, S\~ao Paulo, SP 05508-090, Brazil\\
$^2$ School of Mathematics, Statistics and Physics, Newcastle University, NE1 7RU, UK \\
$^3$Kavli Institute for Particle Astrophysics and Cosmology, Stanford University, Stanford, CA 94305, USA\\
$^4$Departamento de F\'isica, Centro de Ci\^encias Naturais e Exatas, Universidade Federal de Santa Maria, 97105-900 Santa Maria, RS, Brazil
}

\date{Accepted 2023 August 30. Received 2023 August 2; in original form 2023 February 28}
\pubyear{2023}

\begin{document}
\label{firstpage}
\pagerange{\pageref{firstpage}--\pageref{lastpage}}
\maketitle

\begin{abstract}
We present a simple model for low-luminosity active galactic nucleus (LLAGN) feedback through winds produced by a hot accretion flow. The wind carries considerable energy and deposits it on the host galaxy at kiloparsec scales and beyond, heating the galactic gas thereby quenching star formation. Our model predicts that the typical LLAGN can quench more than $10\%$ of star formation in its host galaxy. We find that long-lived LLAGN winds from supermassive black holes (SMBH) with masses $\geq 10^8 M_{\odot}$ and mass accretion rates $\dot{M} > 10^{-3} \medd \ (0.002 \msun / yr)$ %10^{-3}\dot{M}_{\rm Edd}$ c
an prevent gas collapse and significantly quench galactic star formation compared to a scenario without AGN, if the wind persists over 1 Myr. For sustained wind production over timescales of 10 Myr or longer, SMBHs with $10^8 M_{\odot}$ or larger masses have important feedback effects with $\dot{M} > 10^{-4} \medd \ (0.0002 \msun / yr)$.%10^{-4} \dot{M}_{\rm Edd}$. 
\end{abstract}


%%%%%%%%%%%%%%%%%%%%%%%%%%%%%%%%%%%%%%%%%%%%%%%%%%

%%%%%%%%%%%%%%%%% BODY OF PAPER %%%%%%%%%%%%%%%%%%

\section{Introduction}\label{sec:introduction}

% Synonyms: early-type, red sequence, quiescent
Once an early-type galaxy forms, that does not mean it will remain quiescent forever and ever. Early-type galaxies have abundant gas \eg{Binette1994} and should also accrete fresh amounts of it. If all this gas cooled and led to star formation, the global stellar mass density should currently be larger than observations by a factor of a few \citep{Benson2003}. Furthermore, the number of galaxies in the red sequence is steadily growing since the peak epoch of quasars and starbursts \eg{Bell2004, Bundy2006}. This implies that galaxies are still transitioning to quiescence. Taken together, these are evidence for an unceasing feedback process which suppresses star formation in red sequence galaxies and keeps it quenched.

In this work, we explore the possibility that the feedback mechanism keeping these galaxies quiescent is due to winds from accreting supermassive black holes (SMBH) hosted in low-luminosity active galactic nuclei (LLAGN). This idea is quite promising because most SMBH activity in the nearby universe is happening in LLAGNs \eg{Ho2008}. These SMBHs are weakly accreting via radiatively inefficient accretion flows (RIAF; \citealt{Yuan2014}). RIAFs are prone to producing profuse winds \eg{Yuan2015,Almeida2020, Yang2021}. In addition, there is increasing evidence of a new class of early-type galaxies hosting galaxy-scale LLAGN winds from spatially resolved spectroscopy \citep{Cheung2016, Roy2021, Sanchez2021} and radio observations \citep{Roy2018}.

%Motivation
Given the potential importance of AGN winds in quenching star formation at late times, here  we perform an analytical study of LLAGN winds as a feedback mechanism. We build a simplified model of RIAF winds based on the latest results from numerical simulations and analyze how the presence of an LLAGN could impact the gas and stellar content of a galaxy.

RIAF winds are very hot, subrelativistic and non-collimated. They carry considerable energy, with powers up to 1\% of the rest mass energy $\dot{M} c^2$ associated with accretion \citep{Almeida2020}. The kinetic and thermal energy of the ejected wind must be deposited in the environment, and its most plausible fate is depositing its energy in the interstellar medium. By exploring the properties of these winds and their impact on the host galaxy, we tackle the following questions: Are LLAGN powerful enough to quench star-formation in an early-type galaxy? Can LLAGN winds keep a red-and-dead galaxy quiescent?

%Structure
This paper is structured as follows. In section \ref{sec:model}, we present the details of the model. In section \ref{sec:results} we present the results, which include the predicted relation between LLAGN power and star-formation quenching. We compare our results to the literature in section \ref{sec:discussion}. Finally, section \ref{sec:summary} presents a summary and some perspectives.

\section{Model} \label{sec:model}

In order to quantify the effect of LLAGN feedback, we approximated a galaxy as an isothermal sphere of dark matter with a fixed fraction of gas. The wind itself is an expanding sphere. %that cools via bremsstrahlung. 
In the following subsections, we describe our model in more details.

\subsection{Galaxy}

We followed \cite{Silk1998} and modelled the galaxy as an isothermal sphere characterized by a velocity dispersion $\sigma$. Stars dominate the total mass of the galaxy's central region, and only a small fraction is gaseous corresponding to a fraction $f_g \approx 0.05-0.1$ of the total mass. The gas density profile is described as
\begin{equation}
    \rho (R) = \frac{f_{g} \sigma^2}{2\pi G R^2}.
    \label{eq:galaxy-density}
\end{equation}
The total gas mass enclosed in a radius $R$ is
\begin{align}
M_{\rm gas}(R)  & = \int_0^R 4\pi r^2 \rho(r) dr = \frac{2f_g\sigma^2R}{G} \nonumber \\
      & = 9.6 \times 10^{9} f_g \left( \frac{\sigma}{200 \, {\rm km/s}} \right)^2 \left( \frac{R}{1 \, {\rm kpc}} \right) M_{\odot} \label{eq:galaxy-mass}
\end{align}
and is in the form of atomic hydrogen. The gravitational binding energy $E_{\rm gal}$ is 
\begin{equation}
    E_{\rm gal}(R) = \frac{3GM_{\rm total}M_{\rm gas}}{5R} = \frac{6M_G\sigma^2}{5}.   
    \label{eq:galaxy-gravitational-binding-energy}
\end{equation}
Adopting $f_g = 0.05$ and replacing equation \eqref{eq:galaxy-mass} in \eqref{eq:galaxy-gravitational-binding-energy} gives
\begin{equation}
    E_{\rm gal}(R) = 4.5 \times 10^{56} \left( \frac{\sigma}{200 \, {\rm km/s}} \right)^4 \left( \frac{R}{1 \, {\rm kpc}} \right) \text{ erg }.  
    \label{eq:galaxy-energy}
\end{equation}
The system is isothermal with a temperature of $T_{\rm Gal} = 1.5 \times 10^{6} \sigma_{200}^2 \text{ K }$ where $\sigma_{200} \equiv \sigma/200 \, {\rm km/s}$.

\subsection{LLAGN Energy Output}

The LLAGN is able to inject a $\Delta E$ amount of energy into the galaxy via winds given by $\Delta E = L_w \Delta t$. Being $L_w$ the wind power and $\Delta t$ is the LLAGN lifetime. We parameterise the wind power as a fraction of the Eddington luminosity, $L_w = \eta L_{\rm Edd}$. Following \cite{Almeida2020}, the wind power is $\sim 0.1-1$ per cent of the rest-mass energy $\dot{M} c^2$ accreted by the SMBH. Given that for a LLAGN we expect $\dot{M} \lesssim 10^{-3} \dot{M}_{Edd}$ and $L_{\rm Edd} \equiv 0.1 \dot{M}_{\rm Edd}c^2$, we have $\eta \lesssim 10^{-4}$. Thus, in our calculations we assume $\eta = 10^{-4}$ and thereby
\begin{equation}
    \Delta E = 4 \times 10^{56} \left( \frac{\eta}{10^{-4}} \right) \left( \frac{M_{\rm BH}}{10^9 \, {\rm M}_{\odot}} \right) \left( \frac{\Delta t}{1 \, {\rm Myr}} \right) \ \text{erg}.
\label{eq:AGN-energy-output}
\end{equation}

With these considerations, the impact of the AGN on the host galaxy increases trivially with its lifetime and decreases with the distance from the SMBH, as can be seen by taking the ratio of the LLAGN energy output with the galactic gravitational binding energy, 

\begin{equation}
f_{\rm AGN} \equiv \frac{\Delta E}{E_{\rm gal}} = 0.24 \left( \frac{\Delta t}{1 \, {\rm Myr}} \right) \left( \frac{R}{1 \, {\rm kpc}} \right)^{-1} \left( \frac{M_{\rm BH}}{10^9 \, {\rm M}_{\odot}} \right)^{0.22},  \label{eq:AGN-energy-fraction-m-sigma}
\end{equation}

\noindent where we have used the $M-\sigma$ relation from \cite{McConnell2011}, 

\begin{equation}
\left( \frac{M_{\rm BH}}{10^8 \msun} \right) = 1.95 \left( \frac{\sigma}{200 \ km \ s^ {-1}} \right)^{5.12} .
    \label{eq:M-sigma-Mcconnell}
\end{equation}

\noindent As we will see, the LLAGN energy output can be comparable to the galactic gravitational binding energy.

\subsection{Star-formation}

%General Model
Star formation usually occurs in giant molecular clouds (GMC), massive reservoirs of cold gas prone to star formation. 

In our model, we assume that the entirety of the wind kinetic power couples to GMCs and is converted to thermal energy. This approximation amounts to $f_{\rm AGN}$ translating directly into the fractional temperature increase caused by AGN feedback.

We describe the protostellar core mass function as
\begin{equation}
    \frac{dN}{dlnM} = N_0 \Big( \frac{M}{M_0} \Big)^{-\xi} \text{,        } (M \lesssim M_0).
    \label{eq:CMF}
\end{equation}
following \cite{Rosolowsky2005, Dib2008}. Equation \eqref{eq:CMF} gives the distribution of protostellar cores inside GMCs as a function of mass and sizes. We considered in our model dense clouds with $M_0 \lesssim 100 M_\odot$ and $0.3 \leq \xi \leq 2.7$ \citep{Drapatz1984, Rosolowsky2005, Dib2008, Mok2020, Rosolowsky2021}.

%Jeans Mass
Cores able to generate stars are those with masses exceeding the Jeans mass 
\begin{equation}
M_J = 20 M\odot \Big(\frac{T}{10 \ {\rm K}}\Big)^{1.5} \Big(\frac{n}{100 \ {\rm cm}^{-3}}\Big)^{-0.5}.
\label{eq:jeans}
\end{equation} 
Assuming, in our model, a constant external pressure around the cloud and $n=100 \ {\rm cm}^{-3}$, equation \eqref{eq:jeans} simplifies to $M_J = 20 M_\odot (T/10 \ {\rm K})^{2}$.

\section{Results} \label{sec:results}

\subsection{Energetics} 

\begin{figure*}
    \centering
    \includegraphics[width=\linewidth]{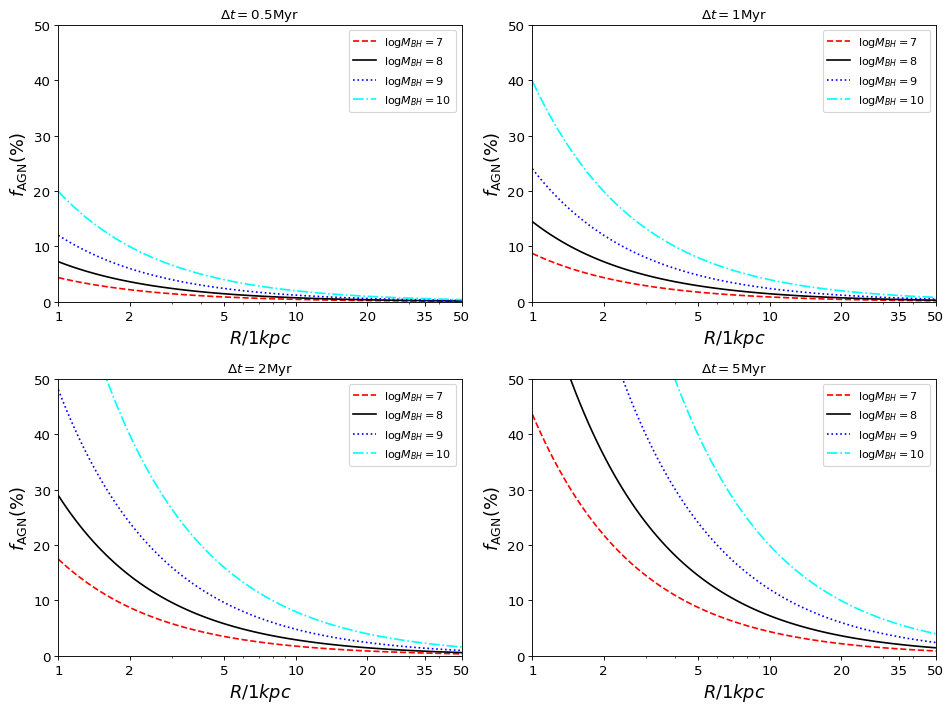}
    \caption{Energy injected by LLAGN winds scaled by the galactic binding energy as a function of distance to the supermassive black hole, based on equation \ref{eq:AGN-energy-fraction-m-sigma}. Different AGN durations and black hole masses are displayed in the different panels, with the mass in solar masses.}    
    \label{fig:deltat-Rw}
\end{figure*}

Figure \ref{fig:deltat-Rw} illustrates the characteristic values of $f_{\rm AGN}$ for a range of AGN timescales and distances. The figure indicates that an LLAGN can inject a significant amount of energy into the inner 10 kpcs of the host galaxy. The effect is more prominent in galaxies with more massive SMBHs. For instance, a galaxy hosting a $10^8 M_\odot$ SMBH can undergo a $10\%$ temperature increase in the innermost 2 kpc in one million years; a $10^{9} M_\odot$ SMBH active over 2 Myr with achieve a heating fraction higher than 50\%. Moreover, if the LLAGN is active for 5 Myr or longer, the galactic heating within 5 kpc will be energetically relevant regardless of the mass.

\subsection{How far does the wind reach?} \label{subsec:WZ-size}

Simulations suggest strong winds coming from RIAFs, with powers reaching up to one percent of the rest mass associated with accreted gas \citep{Almeida2020}. These winds have thermal energies greater than the gravitational binding energy, which means they have enough energy to escape the black hole's gravitational sphere of influence. Nevertheless, the spatial extent of these winds remains an open question. We investigated the wind extension using two different approaches. In the first one, we model the wind as expanding bubble which cools via bremsstrahlung. In the second one, we consider a central heating source and a heat transfer through the gas---here, the wind carries only energy and not mass.

In the first scenario, we computed the distance travelled by the bubble front over the cooling time, $R_{\rm wind} = v t_{\rm cool}$, where we assume $v= 300 \ {\rm km \ s}^{-1}$ \citep{Cheung2016,Almeida2020} and that the density follows $\rho_{\rm wind} \propto r^\alpha$. The resulting expression is 
\begin{equation}
    R_{\rm wind} = \Bigg(\frac{5.9 \times 10^{-5}\eta^{-1} \sigma_{200}^{2} M^\alpha}{10^{7\alpha}\sqrt{1-\alpha}} \Bigg)^{\frac{1}{1+\alpha}} \ {\rm kpc} 
    \label{eq:R-wind-bremsstrahlung}
\end{equation}
where we assume $\eta \sim 10^{-4}$, related to the efficiency of the wind production. This is roughly
\begin{equation}
R_{\rm wind} \gtrsim 
    \begin{dcases}
        3 \ {\rm kpc}, \alpha < -0.1\\
        100 \ {\rm kpc}, \alpha < -0.3\\
    \end{dcases}
\end{equation}
We find that for $\alpha < 0$, the wind can reach distances larger than ten kpc which are beyond the visible size of most galaxies.

For the second case, we numerically solve the one-dimensional radial heat transfer equation for a sphere made of hydrogen with a central heat point source, 
\begin{equation}
    \frac{1}{r^2}\partial_r(r^2 \partial_r T) = \frac{\rho c_P r^2}{\kappa}\partial_tT + Q_{\rm AGN}
    \label{eq:heat-transfer}
\end{equation}
We modelled the AGN impact as a spherical boundary with constant temperature and hotter than the medium. This can be translated as the boundary condition in equation \eqref{eq:AGN-boundary} and initial condition in equation \eqref{eq:T-initial-condition}. For practical reasons, we assumed $r_{\rm AGN} = 0$ since the AGN scales are too small compared to the galaxy.
\begin{align} 
& T(r=r_{\rm AGN}) \leq T_{AGN}\text{,}
\label{eq:AGN-boundary} \\  
& T(t=0, r) = 
    \begin{dcases}
        T_{\rm AGN}, & r \leq r_{\rm AGN}\\
        T_{\rm gal}, & r > r_{\rm AGN}
    \end{dcases}.
\label{eq:T-initial-condition} 
\end{align}
Solving equation \eqref{eq:heat-transfer} and assuming the characteristic values from \cite{Fabian2005} (their equation 4),  we found that the resulting temperature profile follows $T(R) \propto R^{-1}$. This is the same radial dependence as in equation \eqref{eq:AGN-energy-fraction-m-sigma}. After about 5 Myr, even gas at kiloparsec scales will undergo a 20\% temperature increase. For this model $R_{\rm wind}$ is the radius at which $\lim _{r \to R_{\rm wind} } T(r)=T_{\rm gal}$. We find that typically $R_{\rm wind} \gtrsim 1$ kpc.

Both models indicate that winds can get to the galactic outskirts, reaching distances up to kpc. We stress that the multiscale physics of the ISM and its interaction with hot winds is quite complex. We leave the numerical modeling of these phenomena for a future work.

\subsection{Star formation quenching}

%quenching rate
The number of protostellar cores able to collapse and form stars can be calculated using equations \ref{eq:CMF} and \ref{eq:jeans} as 
\begin{equation}
    \mathcal{N}(M\geq M_J) = \int_{M_J}^{M_0} N(M) dM.
    \label{eq:SF-number}
\end{equation}
We use $\mathcal{N}$ to quantify the impact of LLAGN feedback in quenching star formation by computing it in two different ways: $\mathcal{N}_0$ is the number of protostellar cores able to collapse into stars when the AGN effect is not taken into account, whereas $\mathcal{N}_{\rm AGN}$ is the corresponding quantity with the AGN turned on. In particular, we are interested in comparing how much lower $\mathcal{N}_{\rm AGN}$ is compared to $\mathcal{N}_0$ as a function of the main accreting BH parameters: the BH and mass accretion rate. When estimating $\mathcal{N}_0$, we consider a temperature $T_{\rm PC} \sim 10$K and corresponding Jeans mass is denoted by $M_J$ (see equation \eqref{eq:jeans}); for $\mathcal{N}_{\rm AGN}$, we adopt $T_{\rm PC}^{\rm AGN} = (1+f_{\rm AGN})T_{\rm PC}$ as the AGN increase the average temperature and the appropriate Jeans mass is $M_J^{\rm AGN}$.
This implies that $M_J < M_J^{\rm AGN}$. Protostellar cores with masses in the range $M_J < m < M_J^{\rm AGN}$ will suffer gravitational collapse when the impact of the AGN is not considered; they would not if the LLAGN is taken into account. 

We define the fraction of star formation quenched by the LLAGN---the quenching fraction $Q$---as 
\begin{equation}
    Q \equiv 1 - \frac{\mathcal{N}_{\rm AGN}}{\mathcal{N}_0} = 1 - \frac{1 - (M_J/M_0)^{1-\xi}(1+f_{\rm AGN})^{2-2\xi}}{1 - (M_J/M_0)^{1-\xi}}.
    \label{eq:SF-quenching-fraction}
\end{equation}
where $\xi$ is a power-law index and $M_0$ is the mass scale related to the protostellar core mass distribution, (see equation \eqref{eq:CMF}). 
The meaning of $Q$ is the following: in the extreme case when $Q=1$, the entirety star formation is aborted due to AGN feedback; on the other hand, when $Q=0$ there is no quenching at all. Therefore, $Q$ and the star-formation rate are inversely correlated.

\begin{figure}
    \centering
    \includegraphics[width=\linewidth]{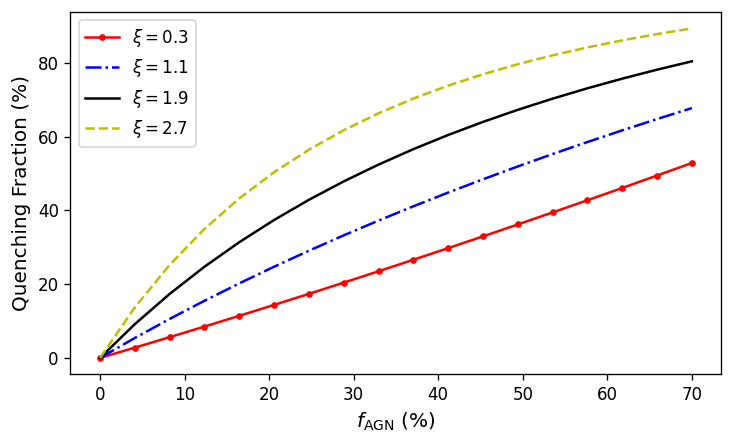}
    \caption{The quenching fraction as a function of the average heating of the region. As the temperature increases, the fraction of shut-down stellar formation sites increases. The different lines represent the different distribution possibilities for the protostellar cores (see equation \eqref{eq:CMF}). }    
    \label{fig:q-xi}
\end{figure}

We plot in figure \ref{fig:q-xi} the relation between star formation quenching and the AGN heating fraction $f_{\rm AGN}$, where we explore the dependence on the parameter $\xi$ (equation \eqref{eq:CMF}). As expected, quenching becomes more pronounced as the amount of energy dumped by the LLAGN increase though this proceeds in a nonlinear fashion.

\begin{figure*}
    \centering
    \includegraphics[width=\linewidth]{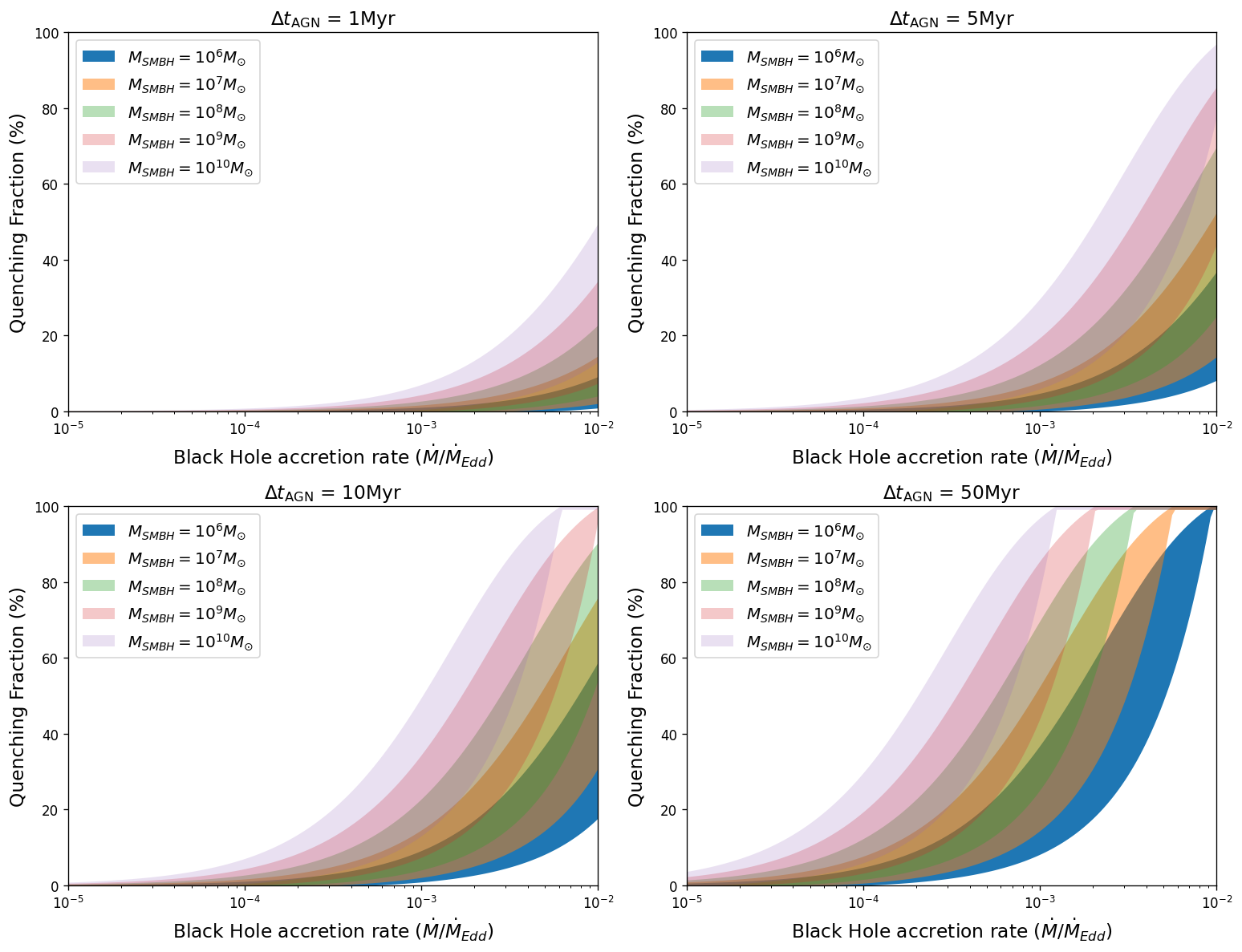}
    \caption{The plot shows the quenching fraction inside a region of 20kpc as a function of the LLAGN accretion rate. The increase in the accretion rate has a significant effect on the gas. Each colour represents a different SMBH mass. We can observe the importance of the system's total mass; the quenching only occurs for the most massive SMBHs. The three different panels refer to the LLAGN activity time $\Delta t$, long-lived LLAGN have a much more substantial impact on the gas temperature and subsequent quenching. The denoted regions represent the different distributions of the protostellar cores (see equation \eqref{eq:CMF}); they are the region delimited by the lines shown in figure \ref{fig:q-xi}.}    
    \label{fig:money-plot}
\end{figure*}

Figure \ref{fig:money-plot} illustrates the dependence of quenching on the SMBH mass accretion rate. Each shaded region with a different color corresponds to a given SMBH mass, with the interior spanning all allowed $\xi$ values assuming $R=20$ kpc (a typical galaxy size). 
%(see equation \ref{eq:CMF} and figure \ref{fig:q-xi}) for different SMBH masses and considering an sphere with 20kpc radius. 
The different panels explore the impact of the duration of the LLAGN activity varying from 1 Myr (upper left panel) to 50 Myr (bottom right panel). For illustration, let's consider a SMBH accreting at the $10^{-3} \dot{M}_{\rm Edd}$ level. If its mass is $10^8 M_\odot$ ($10^9 M_\odot$) and the wind is produced for only 1 Myr, it can quench less than one per cent ($5\%$) of star formation in the host galaxy; now, if the LLAGN is active for 10 Myr, it can quench up $10\%$ ($30\%$); moreover, if it is active for 50Myr, the quenched grows to $40\%$ ($60\%$).

\begin{figure}
    \centering
    \includegraphics[width=\linewidth]{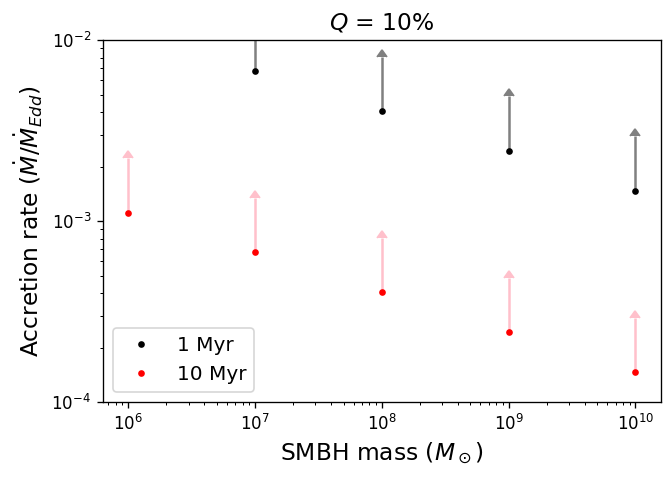}
    \caption{Family of accreting SMBH parameters required to produce  quenching of star formation of at least ten per cent, as a function of BH mass. Black (top) and red (bottom) circles correspond to LLAGN lifetimes of 1 Myr and  10 Myr, respectively. This figure serve as a guide to evaluate whether a LLAGN feedback is effective as a function of black hole mass and accretion rate. }   
    \label{fig:activate}
\end{figure}

Figure \ref{fig:activate} displays the SMBH activation function for effective AGN feedback, as predicted in our calculations. This figure displays the family of accreting SMBH parameters required to produce a ten per cent quenching of star formation, i.e. the combination of mass accretion rates and masses that result in $Q=0.1$. Figure \ref{fig:activate} shows that a $10^8 M_\odot$ or $10^9 M_\odot$ SMBH that experiences an accretion episode lasting 1 Myr with $\dot{M} > 4 \times 10^{-3} \dot{M}_{\rm Edd}$ will be able to abort more than $10\%$ of star formation in its host galaxy. For an accretion episode lasting 10 Myr, a $10^8M_\odot$ SMBH needs $\dot{M} > 4 \times 10^{-4} \dot{M}_{\rm Edd}$ to significantly impact its host galaxy via winds; a $10^9M_\odot$ SMBH needs $\dot{M} > 3 \times 10^{-4} \dot{M}_{\rm Edd}$. 

\begin{figure}
    \centering
    \includegraphics[width=\linewidth]{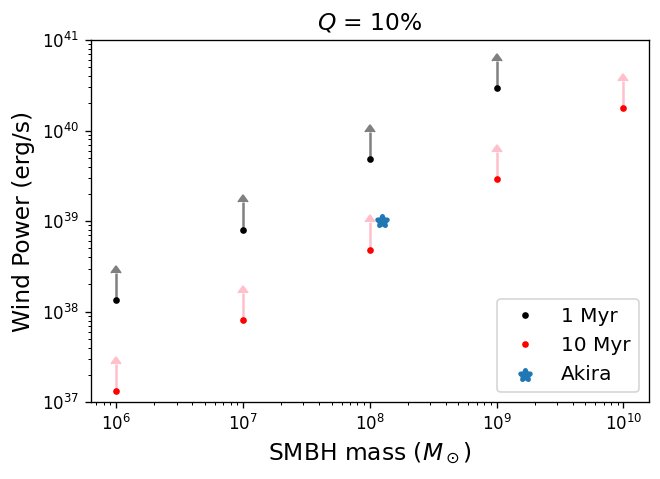}
    \caption{Wind power required to produce quenching of star formation of at least ten per cent. Color scheme is the same of Figure \ref{fig:activate}. The star indicates the values measured for the Akira LLAGN. }   
    \label{fig:power}
\end{figure}

Correspondingly, Figure \ref{fig:power} displays the wind power resulting in effective AGN feedback with $Q \geq 0.1$. Similarly to the story told in Figure \ref{fig:activate}, a $10^8 M_\odot$ SMBH that produces a wind lasting 1 Myr with power larger than $5 \times 10^{39} {\rm erg \ s}^{-1}$ will be able to abort more than $10\%$ of star formation in its host galaxy. For winds lasting 10 Myr, a $10^8M_\odot$ ($10^9M_\odot$) SMBH needs a wind power larger than $4 \times 10^{38}{\rm erg \ s}^{-1}$ ($2 \times 10^{39} {\rm erg \ s}^{-1}$) for effective quenching.

Overall, the LLAGN will only have an impact larger than ten per cent on the host galaxy if it persists for durations longer than 10 Myr, regardless of the SMBH mass. This timescale is one order of magnitude larger than the typical quasar lifetime. Long LLAGN durations are needed in order to significantly suppress star formation.

\section{Discussion} \label{sec:discussion}

Going back to the questions posed at the beginning of this work: Are LLAGN powerful enough to quench star-formation in an early-type galaxy? Can LLAGN winds keep a red-and-dead galaxy quiescent? With our simple models we find that the answer to both questions is yes. The quenching intensity, however, depends on the black hole mass, accretion rate and on the duration of the accretion episode. 

The accretion rate is a crucial parameter in our model. By converting the Eddington units to solar masses per year ($\msun$/yr), we obtain:
\begin{equation}
\medd = 2.23 \left( \frac{M_{\rm SMBH}}{10^8 \msun} \right) \msun \ \text{yr}^{-1}.
    \label{eq:Eddington-rate-to-physical-units}
\end{equation}
For a black hole with a mass of $10^8 \msun$ accreting at $10^{-3} \medd$, the total mass accreted per year is $0.002 \msun/\text{yr}$. In our extreme case of a $10^9 \msun$ black hole accreting at $10^{-2} \medd$, the required amount of gas is approximately $0.2 \msun/\text{yr}$. The mass necessary to trigger our winds is highly feasible, considering the dense nature of the central regions of galaxies. Now, if we consider the total amount of gas needed to sustain these two scenarios for a duration of 1 Myr, we would require $\sim 10^3 \msun$ and $\sim 10^5\msun$, respectively. These values still remain significantly below the mass scales of the supermassive black hole or the stellar mass of the galactic centre.

Let's consider now the particular case of the \quote{Akira} galaxy. 
%Red Geysers
\cite{Cheung2016} reported evidence for winds emerging from the LLAGN in Akira. The authors dubbed this putative class of objects ``red geysers'' \eg{Roy2018}. Our work supports the notion that LLAGN winds can indeed be energetic enough to originate the red geyser phenomenon. \cite{Cheung2016} find that Akira hosts a $10^8 M_\odot$ SMBH currently accreting with $\lambda \equiv L/L_{\rm Edd}=4 \times 10^{-4}$ and that the wind lasts at least 10Myr. This value of $\lambda$ corresponds to $\dot{M}=3\times 10^{-3} \dot{M}_{\rm Edd}$, for a typical RIAF radiative efficiency of $1\%$ \citep{Xie2012}. Our model predicts that LLAGN winds in Akira can reach quenching fractions of about $30\%$ if those accretion rates are sustained over 10 Myr, and potentially much more for longer times. Star formation in the so-called red geyser galaxies can be significantly impacted by winds produced from underfed SMBHs.

%The wind's reach
We explored two different assumptions on the radial expansion of the wind. Both of them indicate that the kinetic and thermal energies can be carried over kiloparsec scales way beyond the SMBH gravitational sphere of influence.

%activity time
An important parameter in our results is the activity time of the LLAGN. If we want to explain the quiescence of the local universe galaxies as the effect of a steady and weak wind from a very faint AGN, this object must be active for a very long time. In figure \ref{fig:money-plot}, we can see in the left panel that only SMBHs with mass $M_{\rm SMBH} \gtrsim 10^9M_\odot$ and $\dot{m}\gtrsim 5 \times 10^{-3}$ can noticeably impact the star-formation in $\Delta t_{\rm AGN} = 1$Myr. However, for a longer time as $\Delta t_{\rm AGN} = 10$Myr, one LLAGN with $\dot{m} \gtrsim 10^{-3}$ and masses $M_{\rm SMBH} \gtrsim 10^8M_\odot$ can turn off more than 50\% the stellar formation sites. The star formation can be severely suppressed if the galaxy inflow can sustain the LLAGN accretion rate for a long enough time.

%metallicity and reddening
One limitation of our model is that we are unable to give more details on specific types of stellar populations arising after quenching by the LLAGN winds. Modeling the vast dynamical range and the nonlinear physics involved in star formation is complex problem and outside the scope of this work -- a simulation of effect feedback for an elliptical galaxy treated in much more detail can be seen in \cite{Yuan2018}. One broad brush consequence of the suppression of star formation is that there will be a smaller amount of heavy elements being spewed out throughout the galaxy. Thus, galaxies under the influence of LLAGN feedback will have smaller metallicities. At the same time, and for the same reasons, we expect a smaller number of younger stars, so LLAGN winds tend to redden the host galaxy.

Our model assumes a smooth wind that interacts with molecular clouds, heating them up over Myr timescales. In a more realistic setting, outflows likely strip gas clouds. The ensuing cloud mass decrease would further boosting the quenching fraction to higher values than we reported in figure \ref{fig:money-plot}. This possibility remains to be investigated in the future.

Another aspect worth discussing is the feedback impact on the black hole mass supply itself, i.e. does the LLAGN feedback reduces significantly $\dot{M}$? According to \cite{Bu2019} at first feedback inhibits $\dot{M}$, which also leads to a reduction in the wind strength; feedback then becomes less important and accretion eventually is reestablished to its previous levels. \cite{Bu2019} find that this cycle occurs on timescales orders of magnitude shorter than those we consider here. For instance, when the $\dot{M}$ valus reported on Bu \& Yang are averaged over timescales longer than $10^5$ years, the accretion rate is well-behaved (cf. their Figure 2c). This supports our assumption that $\dot{M}$ does not vary much over timescales longer than 1 Myr.

Furthermore, another argument can be made that $\dot{M}$ is not severely impacted by the LLAGN feedback. Keeping the assumption of a spherically symmetric, stationary accretion flow, we consider the outflow depositing energy at the Bondi radius $r_B$. By solving the one-dimensional fluid conservation equations with appropriate boundary conditions (cf. appendix in \citealt{DiMatteo2003}), the Bondi radius is modified due to the outflow heating as 
\begin{equation}
r_B=\frac{G M}{2 c_s^2}-\frac{2}{3} \frac{H r_B^2}{c_s^3},
\end{equation}
where the first term on the right side is the usual Bondi radius and the second term represents the heating effect; $H$ is the heating rate in units of erg s$^{-1} \ {\rm g}^{-1}$ given by $H = L_w/M_g$ where $M_g \sim 4\pi r_B^3 \rho/3 $; $\rho$ and $c_s$ are evaluated at $r_B$. Since $L_w = \eta \dot{M} c^2$ and assuming $\dot{M} = \epsilon \dot{M}_B$, i.e. the BH accretion rate is a small fraction of the Bondi rate per ADIOS scenarios, we find to first order:
\begin{equation}
r_B \approx \frac{G M}{2 c_s^2}\left[1-2 \eta \epsilon \left( \frac{c}{c_s} \right)^2 \right],
\end{equation}
being $c$ is the speed of light. Assuming typical wind feedback values of $\eta \sim 0.01$ \eg{Almeida2020}, $\epsilon \sim 0.01$ \eg{Yuan2014} and $c/c_s \sim 10^3$, we find that $r_B$ could be reduced by $\sim 10\%$ due to the LLAGN feedback. Given that $\dot{M} \propto r_B^2$, this implies that the accretion rate is affected only at the per cent level. This is not enough to impact our assumption of a constant $\dot{M}$ in the feedback calculations of the paper. 

The focus of our model is on the evolutionary processes of a single isolated galaxy. The consideration of events such as mergers or interactions with other galaxies, which can result in material transfer between them, falls outside the scope of this study. We adopt a fixed gas mass fraction ranging from $0.05$ to $0.1$ of the total system mass. While we recognise that even isolated galaxies can experience gas infall from the surrounding environment over time, the typical infall rate at redshift $z \sim 0$ is generally not significantly larger than $1 \msun/\text{yr}$ \citep{Sancisi2008, Molla2016}. This rate has a minimal impact on the assumed value for the gas mass fraction within the timescales considered in our study. It is worth noting that gas infall plays a more substantial role on much larger scales than those investigated in this work.

\section{Summary} \label{sec:summary}

The main conclusions of our investigation can be summarised as follows:

(i) Low-luminosity active galactic nuclei can have important feedback effects in their host galaxies by quenching star formation. This occurs via winds emerging from the hot accretion flow which are able to heat up protostellar clouds and prevent them from gravitationally collapsing. 

(ii) The relevance of star formation quenching by LLAGN feedback is a function of the SMBH mass, mass accretion rate and the duration of the accretion episodes. In general, quenching is only relevant for accretion lasting longer than 1 Myr.

(iii) For an accretion episode lasting 1 Myr, a $10^8 M_\odot$ or $10^9 M_\odot$ SMBH needs $\dot{M} \gtrsim 10^{-3} \dot{M}_{\rm Edd}$ to abort more than $10\%$ of star formation.

(iv) For an accretion episode lasting 10 Myr, a $10^8M_\odot$ or $10^9 M_\odot$ SMBH needs $\dot{M} \gtrsim 10^{-4} \dot{M}_{\rm Edd}$ to significantly impact its host galaxy via winds. 

(v) LLAGN winds can reach kiloparsec scales, and beyond.

Our model is subject to the limitations of our assumptions, mainly:  the assumption of a spherical isothermal galaxy, steady state, lack of details on the treatment of the interstellar medium and the wind physics. Despite these idealizations, we hope that our calculations can offer insights on the galaxy-SMBH coevolution.

In conclusion, our model demonstrates that feedback via winds from LLAGNs is an important suppressor of star formation in red sequence galaxies. LLAGNs, despite their low Eddington ratios, will keep a red-and-dead galaxy quiescent at late times. Winds from underfed SMBHs offer a third mode of AGN feedback, in addition to the quasar or radiative mode relevant at the peak of galaxy mergers, and the radio or jet mode relevant for radio galaxies in galaxy clusters.

\section*{Acknowledgements}
We acknowledge useful discussions with Raniere de Menezes, Paula R. T. Coelho, Stephane V. Werner, Feng Yuan, Roger Blandford, Thaisa Storchi-Bergmann, and Ciriaco Goddi. This work was supported by FAPESP (Funda\c{c}\~ao de Amparo \`a Pesquisa do Estado de S\~ao Paulo) under grants 2017/01461-2, 2019/10054-7 and 2022/10460-8. We acknowledge funding from an United Kingdom Research and Innovation grant (code: MR/V022830/1). RN acknowledges a Bolsa de Produtividade from Conselho Nacional de Desenvolvimento Cient\'ifico e Tecnol\'ogico. RAR acknowledges support from CNPQ (400944/2023-5 \& 404238/2021-1) and FAPERGS (21/2551-0002018-0).

We used \texttt{Python} \citep{python2007, python2011} to produce all scientific results and plots of this paper, including several packages such as \texttt{NumPy} \citep{numpy}, \texttt{SciPy} \citep{scipy}, and \texttt{Matplotlib} \citep{matplotlib}.

\section*{Data Availability}
The data underlying this article will be shared on reasonable request to the corresponding author.

%%%%%%%%%%%%%%%%%%%%%%%%%%%%%%%%%%%%%%%%%%%%%%%%%%

\bibliographystyle{mnras}
\bibliography{refs}
% Don't change these lines
\bsp	% typesetting comment
\label{lastpage}
\end{document}